\documentclass[prl, aps, twocolumn, fleqn]{revtex4}
\usepackage{amssymb}
\usepackage{amsmath,amsthm}
\usepackage{graphicx}
\begin{document}

\title{Particle kinematics in a dilute, 3-dimensional,  vibration-fluidized granular medium}

\author{Hong-Qiang Wang}
\email{hqwang@physics.umass.edu}

\author{Klebert Feitosa}
\author{Narayanan Menon}
\email{menon@physics.umass.edu}
\affiliation{%
Department of Physics, University of Massachusetts, Amherst,
Massachusetts 01003-3720
}%

\date{\today}

\begin{abstract}
We report an experimental study of particle kinematics in a
3-dimensional system of inelastic spheres fluidized by intense
vibration. The motion of particles in the interior of the medium is
tracked by high speed video imaging, yielding a spatially-resolved
measurement of the velocity distribution. The distribution is wider
than a Gaussian and broadens continuously with increasing volume
fraction. The deviations from a Gaussian distribution for this
boundary-driven system are different in sign and larger in magnitude
than predictions for homogeneously driven systems. We also find
correlations between velocity components which  grow with increasing
volume fraction.


\end{abstract}


\maketitle





 The distribution of particle velocities is a
fundamental descriptor of the statistics of a particulate system. In
thermal equilibrium, this distribution is always a Boltzmann
distribution, but that is typically not the case for a
nonequilibrium steady state. An archetype of such a steady state is
the inelastic gas, a system of particles that interact by
dissipative contact forces, with the energy lost being compensated
by a driving mechanism. Since the inelastic gas is both of
fundamental interest and closely related to technologically
important granular media, it has been the subject of much recent
experimental, simulational and theoretical activity\cite{Barrat05}.
Most experiments of inelastic gases have focused on 2-dimensional
(2D) systems; in this article, we present the first detailed
measurement of the velocity distribution in a fully 3-dimensional
(3D) gas of inelastic grains.

Inelastic steady states may be broadly divided into systems driven
from the boundaries and those that are driven homogeneously in the
bulk. Homogeneously heated granular gases are represented
theoretically by inelastically colliding particles energized by
random, spatially homogeneous, uncorrelated boosts of energy between
collisions. It has been shown \cite{Noije98} that the high-velocity
tail of the velocity distribution for this model is of the form
$P(c) \propto exp(-Ac^{3/2})$ where $c=v/\sigma$ is a velocity
component, $v$ normalized by its r.m.s value $\sigma$. The velocity
distribution for low velocities was computed by finding perturbative
solutions to the Boltzmann equation in an expansion in Sonine
polynomials around a Gaussian \cite{Goldshtein95, Noije98}. In both
limits, the results depend on inelasticity but not on volume
fraction, $\phi$.  These results have been confirmed by
simulation\cite{Brey96, Monterato00, Moon01}. The closest
experimental analogues for this model are 2D monolayers of particles
fluidized by a vibrated base\cite{Losert99} where velocity
distributions depend strongly on the nature of the
base\cite{Prevost02}; for a sufficiently rough base
\cite{Shattuck07}, the Sonine expansion describes the distribution
satisfactorily. There are no 3D realizations of the homogeneously
heated granular gas with random forcing in the bulk of the medium.

Much less theoretical effort has been directed toward the more
naturally prevalent boundary-driven system. On the other hand,
several experiments have studied 2D systems driven by vibration
\cite{Menon00, Kudrolli00} and electrostatic driving
\cite{Olafsen02}. Some experiments \cite{Menon00, Olafsen02}show the
entire distribution of velocity fluctuations can be described by the
functional form $P(c)\!\propto\!exp(-Ac^{\beta})$ with
$\beta\approx\frac{3}{2}$ over a broad range of number density and
inelasticity. The relationship of these experimental results to the
prediction of \cite{Noije98} are unclear since simulations of the
homogeneously heated gas \cite{Barrat03} show that the asymptotic
high-velocity behavior only sets in extremely deep in the tail, too
rare to be experimentally detectable. Furthermore, the measured
distribution differs between experiments, possibly because these
quasi-2D experiments are sensitive to the specifics of the
confinement in the thin dimension \cite{Menon00, Olafsen04,
vanZon04} or to the substrate on which they move \cite{Kudrolli00}.

Three-dimensional steady states have been much less studied due to
the challenges of tracking particles in the interior of a system.
Two techniques that have been employed for 3D vibrated systems are
positron emission tracking \cite{Wildman01} of tracer particles, and
nuclear magnetic resonance imaging \cite{Huan04}. These studies
focused on spatial profiles of temperature and number density,
although a non-Gaussian velocity distribution was reported by the
NMR technique \cite{Huan04}. Simulations show \cite{Brey03,
Barrat04, Moon04, Zippelius04, MacKintosh} that the velocity
distribution evolves continuously from a nearly-Gaussian
distribution to $P(c)\!\sim\! exp(-Ac^\beta)$. There are no direct
predictions for boundary-driven systems, but models
\cite{MacKintosh, Machta05} which vary $q$, the ratio of the
frequency of particle collisions to the frequency of heating events,
give some intuition on the passage from the homogeneously heated
($q\!\approx\!1$) to the boundary-driven ($q\!\gg\!1$) case.

In this article we describe experiments using high-speed video
imaging to directly locate and track particles in the bulk of a 3D
vibration-fluidized steady state. We report kinetic temperature
profiles, the form of velocity distribution, and comparisons with
available predictions.

\begin{figure}
 \centering
 \includegraphics[width=.4\textwidth, bb=11 11 326 288]{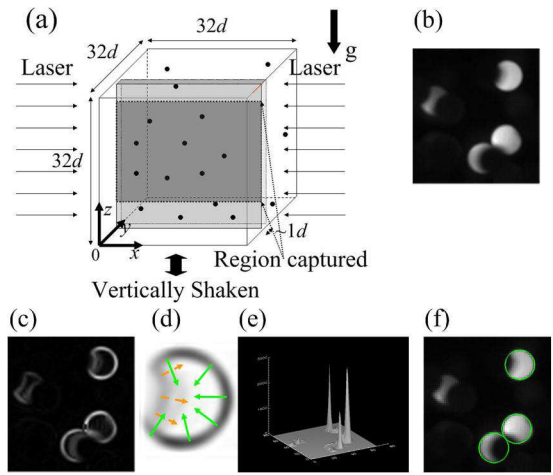}
 \caption{\label{demo} (a) Sketch of the experimental setup. A cubical cell is driven
 vertically by a shaker. The shaded rectangle is a
 plane of thickness $1d$ illuminated by laser sheets from both sides. The darker shading shows the
 area imaged by the camera. (b) Part of a video frame. (c) Edge detection performed on
this frame. (d) Illustration of the rays emanating from the edge,
along the intensity gradient, for the top right particle in (b). The
solid rays converge toward the center and dashed rays diverge. (e)
Surface plot for these accumulated gradient rays. (f) Located
particle positions. For the value of the threshold applied to the
peaks in (e), the fourth particle in the frame is not located. }
\end{figure}

As shown in Fig.\ref{demo}(a), a cubical cell with  acrylic vertical
walls is sinusoidally vibrated in the vertical direction by an
electromechanical shaker (Ling Dynamics V456). The vibration
frequency used in the experiment ranges from 50Hz to 80Hz,  velocity
amplitudes $V_{0}$ from 2.3 m/s to 3.7 m/s, and accelerations from
90$g$ to 190$g$. The bottom and top walls are rough glass plates
that provide a low-inelasticity surface but also randomise the
direction of collisional momentum transfer.
We use delrin spheres of diameter $d=1.560\pm0.003$mm and an average
normal coefficient of restitution $\varepsilon=0.92 \pm 0.21$,
experimentally determined from particle collisions. The error bar
reflects the dependence of $\varepsilon$ on impact parameter,
relative particle velocity, and spin. The side of the cell is
$51.2$mm $\approx 32d$ and is illuminated by light sheets produced
by expanding beams from laser diodes (Thorlabs ML101J8) with a
cylindrical lens. The thickness of the sheet in the $y$-direction is
$\approx\!1d$, and its $y$-position can be varied, allowing us to
study particle motions in $x$-$z$ planes at varying depth from the
front wall. A Phantom v7 camera images the $x$-$z$ plane selected by
the light sheet, at 5,000 frames/second, with a resolution of
$640\times480$ pixels. The field of view is a rectangle of width
$\approx 30d$ and height $20d$, centred on the middle of the cell.

The difficulty in locating particles
 stems from two distinct eclipsing effects. First, particles in the
 middle of the cell are less likely to be
illuminated because the light-sheet is obstructed by particles in
the light path. The delrin spheres are homogeneously illuminated;
any laser light incident on a sphere is scattered through its
volume. However, when the light-sheet is at some depth from the
front plane, illuminated spheres can be partially eclipsed by
particles in front of them. We focus on velocity statistics rather
than number density distributions since the detection probability of
a particle depends on these two effects. A video frame with examples
of partially eclipsed particles is shown in Fig.\ref{demo}b. To
locate particles, we exploit the fact that the convex illuminated
edges of the images are circular arcs of known radius. We find
illuminated edges (Fig.\ref{demo}c) and draw rays from all edge
points in the direction of the local intensity gradient. For points
on convex illuminated edges, these rays converge to the centre,
while concave edges produce rays that diverge (Fig.\ref{demo}d).
Local maxima for accumulations of rays (Fig.\ref{demo}e) above a
cut-off value are candidates for particle centres; objects with an
insufficient length of illuminated perimeter are rejected. Particle
centres are determined with subpixel resolution by minimising the
squared distance to the gradient rays. Eclipsing produces no
systematic bias in locating particle centres.

\begin{figure}
 \centering
 \includegraphics[width=.4\textwidth, bb=0 0 540 405]{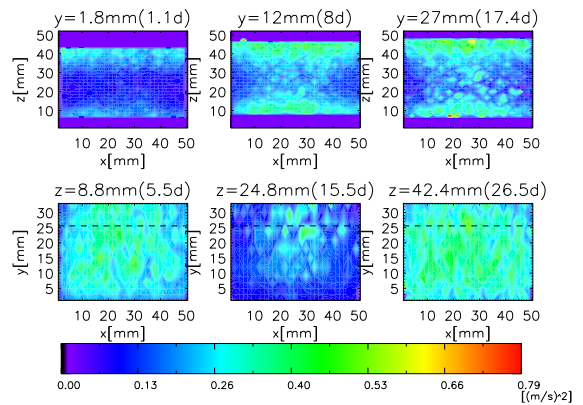}
 \caption{\label{tmap} Distribution of the kinetic temperature at planes of various $y$(top row)
 and $z$(lower row). The dashed line shows the midplane of the cell. The driving frequency is 60Hz, $\Gamma=110g$ and volume fraction 5.1\%.}
\end{figure}

The kinetic temperature
$T=\frac{1}{2}(\left<v_{x}^{2}\right>+\left<v_{z}^{2}\right>)$ is
measured at several depths, $y$, between the front plane and a
little beyond the middle of the cell. As shown in Fig.\ref{tmap}
(top row), in any given $x$-$z$ plane, $T$ is higher near the top
and bottom wall, and lower in the middle. Comparing $x$-$z$ planes
at different depths $y$, we see that $T$ goes up in the middle of
the cell, revealing the dissipative effect of the front vertical
wall. The data are then reorganized to show $T(x,y)$ at a few
heights, $z$ (Fig.\ref{tmap} bottom row). Within each plane, $T$ is
lower close to the vertical walls, again due to the dissipation at
the walls. The length scale over which the wall dissipation
manifests itself is comparable to the mean free path, and grows at
lower $\phi$.

\begin{figure}
 \centering
 \includegraphics[width=.38\textwidth, bb=72 80 514 360]{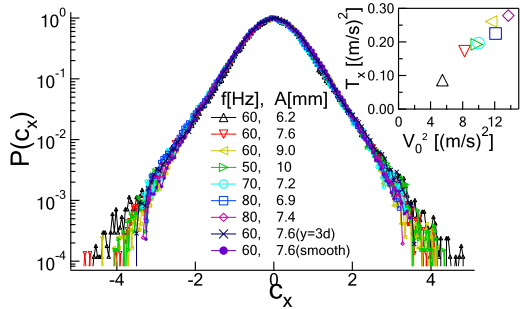}
 \caption{\label{tdriving} $P(c_x)$ for $\phi=5.1\%$
 for various driving frequencies and amplitudes, sampled at $y=8d$. One
 data set taken with smooth top and bottom plates, and another set
 taken at $y=3d$ are included for comparison. The inset shows the average kinetic temperature of the sampled
 region as a function of $V_{0}^{2}$, where $V_0$ is the
 amplitude of the driving velocity.}
\end{figure}

We now turn to the entire distribution of the velocity fluctuations.
The velocity fluctuations are anisotropic; we concentrate on $v_x$,
the horizontal component, perpendicular to the driving direction.
The distribution $P(c_x)$ of $c_x$, the normalized horizontal
velocity $c_x=v_x/(2T_{x})^{1/2}$ where $T_x=\left<v_{x}^2\right>$,
shows weak dependence on position very close to the walls of the
cell (as also seen in simulations \cite{Barrat04, Moon04}).
Therefore, we report data at $y=8d$ which is far from the front
wall, and yet not so deep in the cell where the velocity statistics
are greatly diminished by eclipsing effects. The dependence of
$P(c_x)$ on different driving frequencies and amplitude is shown in
Fig.\ref{tdriving}. As the overall temperature is changed by a
factor of 3, we observe no systematic changes in $P(c_x)$. In
experiments on 2D monolayers \cite{Shattuck07, Prevost02} it was
noted that velocity distribution depended on the smoothness of the
driving surface, as might be expected when interparticle collisions
and heating events occur with comparable frequency. To test whether
the influence of the boundary persists into the interior, we replace
the rough glass plates by smooth delrin plates and do not observe
any change in $P(c_x)$. This suggests that the observed statistics
are a consequence of inter-particle collisions, and are insensitive
to the details of the driving surface.

\begin{figure}
 \centering
 \includegraphics[width=.38\textwidth, bb=75 75 524 402]{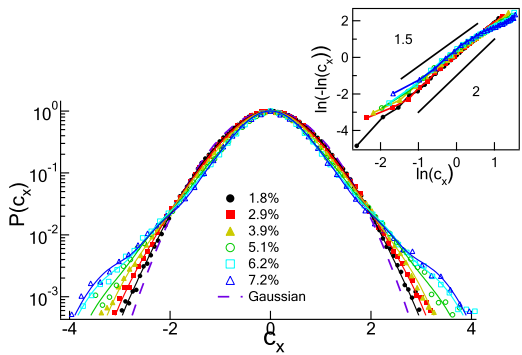}
 \caption{\label{tpv} $P(c_x)$, different volume fractions ranging
 from 1.8\% to 7.2\%. The smooth lines connecting symbols
  are a guide to the eye. The inset shows $ln(-ln(c_x))$ vs $ln(c_x)$.  }
\end{figure}

Having established that distribution of horizontal velocities is
insensitive to $T$, to the driving surface, and to location within
the cell, we now discuss the dependence of $P(c_x)$ on the number
density of particles. We have studied six volume fractions $\phi$
ranging from 1.8 to 7.2\%. The lower limit on $\phi$ is chosen so
that the mean free path is still much smaller than the cell
dimensions, and the upper limit is constrained by poor detection
statistics at high volume fractions. Deviations from a Gaussian are
apparent even at the lowest $\phi$. With increasing $\phi$, the
tails of $P(c_x)$ get broader. Thus the velocity distribution varies
continuously with density unlike in some 2D experiments
\cite{Menon00, Olafsen04} where $P(c_x)$ is unchanged over a broad
range of $\phi$. This is also unlike predictions for the
homogeneously driven or cooled state \cite{Noije98} where $P(c)$ is
independent of $\phi$.

The high-velocity tail of $P(c_x)$ cannot be described  by the form
$exp(-c_{x}^{\beta})$: as shown in the inset of Fig.\ref{tpv}, a
plot of $ln(-ln P(c_x))$ against $ln(c_x)$, shows curvature, whereas
in the equivalent 2D experiment\cite{Menon00} we observed a straight
line with a slope $\beta=1.55\pm0.1$. The statistics in the
experiment only capture the tail up to 4 decades below the peak, and
leave open the possibility that this could be the asymptotic form of
the distribution at large $c$. However, for any realistic
description of grain dynamics, even rarer fluctuations are probably
irrelevant. Earlier simulations\cite{Brey03, Barrat04, Moon04,
Zippelius04, MacKintosh} have found density-dependent velocity
distributions, however, this is the first experimental study in 3D
to observe this effect.

\begin{figure}
 \centering
 \includegraphics[width=.4\textwidth, bb=80 82 512 508]{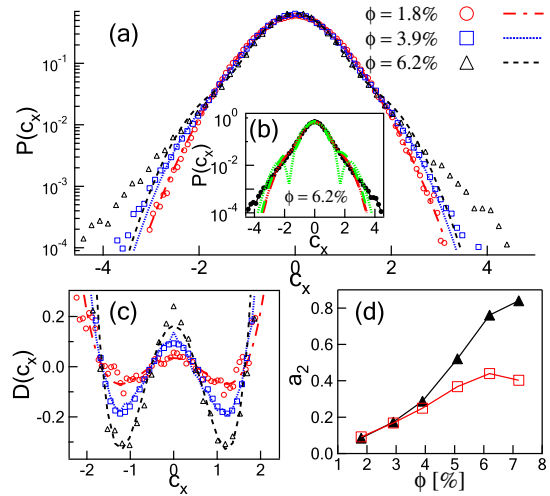}
 \caption{\label{ta2} (a)$P(c_x)$ at volume
 fractions $\phi = 1.8, 3.9$ and $6.2\%$ fitted to second-order Sonine polynomials
 with $a_2$ treated as a free parameter.
 (b)$P(c_x)$ at volume fraction $\phi = 6.2\%$ with best fit second- (red dash line, $a_2 = 0.440$)
 and third-order (red dotted line, $a_2$ = 0.455, $a_3$ = -0.0287)  Sonine forms.
 The green dashed and dotted lines show the 2nd and 3rd order Sonine
 forms with the coefficients $a_2$ and $a_3$ computed from the measured cumulants.
 (c)The deviation function $D(c_x)$ at small $c_x$ for the above volume fractions.
 (d)The best fit value of $a_2$ ($\blacktriangle$) compared with $a_2$ ($\square$) computed from $\left<c^{4}\right>$. }
\end{figure}

In the absence of predictions for the a boundary-driven system, we
compare $P(c)$ to predictions for a homogeneously heated inelastic
gas, where at small velocities, the deviations $D(c)$ from the
gaussian distribution, $\Phi(c)$, have been perturbatively
calculated \cite{Goldshtein95, Noije98} as an expansion in Sonine
polynomials $S_p(c^2)$ :
\begin{equation}
  \label{eq:Sonexp}
  P(c)= \Phi(c)\left[1+D(c)\right] = \Phi(c)\left[ 1
    + \sum_{p=1}^{\infty} a_p S_p\left(c^2\right) \right]  \,.
\end{equation}
The first two polynomials are
$S_1(c^2) = -c^2+\frac{1}{2} \hat{d}$ and $S_2(c^2) =\frac{1}{2} c^4
- \frac{1}{2}(\hat{d}+2)c^2 +\frac{1}{8}
\hat{d}(\hat{d}+2)$ 
where $\hat{d}$ is the dimensionality. The coefficients $a_p$ are
given in terms of the moments of the distribution, $P(c)$:
$    a_1=0, 
    a_2=4\left<c^4\right>/(\hat{d}(\hat{d}+2))-1 $. 
Predictions \cite{Noije98} for the dependence of $a_2$ on the
restitution coefficient for the homogeneously heated and cooling
states have been validated by both DSMC and event-driven
simulations\cite{Brey96, Monterato00, Moon01}. In Fig.\ref{ta2} we
test whether the Sonine expansion is a good description of $P(c_x)$
in our boundary driven system by fitting the data for varying volume
fraction to a second-order Sonine expansion with $a_2$ as a free
parameter. 

 Fig.\ref{ta2}(a)-(c) show that the
second-order Sonine correction works best at the lowest volume
fraction and the range over which $D(c)$, the deviation from a
gaussian, is well-fit diminishes at higher volume fractions. As in
the quasi-2D experiment of \cite{Shattuck07} which models a
homogeneously heated gas; the quality of the fit is reasonable for
$c \lesssim 2$.A third-order Sonine term does not improve the fit,
as shown in Fig.\ref{ta2}(b).

Apart from the dependence on $\phi$ of the fit parameter $a_2$, we
also note that $a_2$ is opposite in sign, and much larger in
magnitude than that found in the homogeneously heated or cooled
states for the same nominal restitution coefficient. Furthermore, as
shown in Fig.\ref{ta2}(d) the best-fit value of $a_2$ disagrees with
the value directly calculated from the 4th cumulant,
$\left<c^{4}\right>$, raising the possibility that the gaussian
reference state may not be appropriate for a boundary-driven system.
To our knowledge, the only predictions for P(c) in a boundary-driven
system were made in \cite{Risso02}, where the 4th cumulant is
treated as an independent hydrodynamic field. They find
$\left<c^{4}\right>$ shows a density dependence qualitatively like
ours, but with much lower magnitude than we measure.

\begin{figure}
 \centering
 \includegraphics[width=.28 \textwidth, bb=86 86 483 385, keepaspectratio]{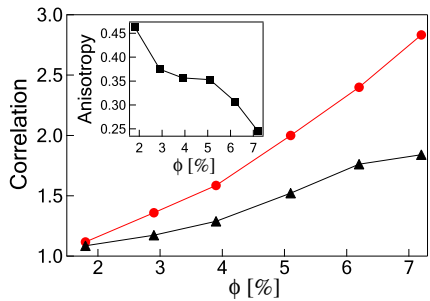}
 \caption{\label{tcorre} The correlation of $v_x$ and $v_z$ ($\bullet$), and
 the correlation implicit in the Sonine formula($\blacktriangle$): $1+a_2$. The inset shows the
 anisotropy, $(v_{z}^{2}-v_{x}^{2})/(v_{x}^{2}+v_{z}^{2})$.}
\end{figure}

A general consideration regarding $P(c)$ is that if a velocity
distribution is isotropic and if the velocity components are
uncorrelated, it must be a gaussian distribution. A non-gaussian
distribution implies that one or both of these assumptions is
invalid. For low volume fractions, the velocity fluctuations are
anisotropic, but they become more isotropic at higher densities.
This trend is shown in the inset to Fig.\ref{tcorre} where the
anisotropy is quantified by
$2\left<v_{z}^2-v_{x}^{2}\right>/\left<v_{z}^2+v_{x}^{2}\right>$.
However, since the velocity distribution does not tend towards a
Gaussian at large $\phi$, the velocity components must be correlated
as shown in Fig.\ref{tcorre}. Indeed, the correlation between
velocity components
$\left<v_{x}^{2}v_{z}^2\right>/\left<v_{x}^{2}\right>\left<v_{z}^2\right>$
grows with volume fraction $\phi$.

Our measurements of the 3D particle kinematics in the interior of a
vibration-fluidized granular medium thus reveal a non-Gaussian
velocity distribution that is insensitive to conditions at the
driving surface. The shape of the distribution evolves continuously
with volume fraction; the functional form differs markedly from the
homogeneously heated state, thus emphasizing the need for
theoretical development for boundary-driven systems.

We are grateful for support through NASA NNC05AA35A and
NSF-DMR0606216, and to R. Soto, M.D. Shattuck, J.L. Machta for
valuable comments. 

\end{document}